\documentclass[11pt,a4paper]{article}
\usepackage{amsthm,amssymb,epsfig,latexsym}
\newcommand{\be}[1]{\begin{equation}\label{#1}}
\newcommand{\ba}[1]{\begin{eqnarray}\label{#1}}
\newcommand{\ee}{\end{equation}}
\newcommand{\ea}{\end{eqnarray}}
\newcommand{\non}{\nonumber\\\rule{0pt}{30pt}}

\newcommand{\num}{\\\rule{0pt}{20pt}}

\newcommand{\dis}{\displaystyle}
\newcommand{\eq}[1]{(\ref{#1})}

\makeatletter
\@addtoreset{equation}{section}
\makeatother

\begin{document}
\begin{flushright}
LPENSL-TH-10/02\\
\end{flushright}
\par \vskip .1in \noindent
\vspace{24pt}
\begin{center}
\begin{LARGE}
{\bf Large distance asymptotic behavior of the emptiness formation
probability of the $XXZ$ spin-$\textstyle{\frac{1}{2}}$ Heisenberg
chain}
\end{LARGE}

\vspace{50pt}

\begin{large}
{\bf N.~Kitanine}\footnote[1]{Graduate School of Mathematical
Sciences, University of Tokyo, Japan,
kitanine@ms.u-tokyo.ac.jp\par
\hspace{2mm} On leave of absence from Steklov Institute at
St. Petersburg, Russia},~~
{\bf J.~M.~Maillet}\footnote[2]{ Laboratoire de Physique, UMR 5672 du CNRS,
ENS Lyon,  France,
 maillet@ens-lyon.fr},~~
{\bf N.~A.~Slavnov}\footnote[3]{ Steklov Mathematical Institute,
Moscow, Russia, nslavnov@mi.ras.ru},~~
{\bf V.~Terras}\footnote[4]{Department of Physics and Astronomy,
Rutgers University, USA, vterras@physics.rutgers.edu \par
\hspace{2mm} On leave of absence from LPMT, UMR 5825 du CNRS,
Montpellier, France}
\end{large}

\vspace{70pt}

\centerline{\bf Abstract} \vspace{1cm}
\parbox{12cm}{\small Using its  multiple integral representation,
we compute the large distance asymptotic behavior of the emptiness
formation probability of the $XXZ$ spin-$\frac{1}{2}$ Heisenberg
chain in the massless regime.}
\end{center}
\vspace{30pt}

\newpage

\section{Emptiness formation probability at
large distance}
The Hamiltonian of the $XXZ$ spin-$1 \over 2$
Heisenberg chain is given by
\be{IHamXXZ} H=\sum_{m=1}^{M}\left(
\sigma^x_{m}\sigma^x_{m+1}+\sigma^y_{m}\sigma^y_{m+1}
+\Delta(\sigma^z_{m}\sigma^z_{m+1}-1)\right).
\ee
Here $\Delta$ is the anisotropy parameter, $\sigma^{x,y,z}_{m}$
denote the usual Pauli matrices acting on the quantum space at
site $m$ of the chain. The emptiness formation probability
$\tau(m)$ (the probability to find in the ground state a
ferromagnetic string of length $m$) is defined as the following
expectation value
\be{EMPtau}
\tau(m)=\langle\psi_g|\prod_{k=1}^m\frac{1-\sigma_k^z}2
|\psi_g\rangle,
\ee
where $|\psi_g\rangle$ denotes the normalized ground state. In the
thermodynamic limit ($M\to\infty$), this quantity can be expressed
as a multiple integral with $m$ integrations
\cite{JimMMN92,JimM96,JimML95,KitMT99,KitMT00}. Recently, in the
article \cite{KitMST02a}, a new multiple integral representation
for $\tau(m)$ was obtained. It leads in a direct way to the known
answer at the free fermion point $\Delta = 0$ \cite{KitMST02c}, in
particular using a saddle point method, and to its first exact
determination outside the free fermion point, namely at $\Delta =
{\frac{1}{2}}$ \cite{KitMST02b}.

The purpose of this letter is to present the evaluation of the
asymptotic behavior of $\tau(m)$ at large distance $m$, in the
massless regime $-1<\Delta<1$, via the saddle point method. We
find
\be{asympt}
\lim_{m\to\infty}\frac{\log\tau(m)}{m^2}=
\log\frac\pi\zeta+
\frac{1}{2}\int\limits_{\mathbb{R}-i0}
\frac{d\omega}{\omega}\frac
{\sinh\frac{\omega}{2}(\pi-\zeta)\cosh^2\frac{\omega\zeta}{2}}
{\sinh\frac{\pi\omega}{2}\sinh\frac{\omega\zeta}{2}
\cosh\omega\zeta},
\ee
where $\cos\zeta=\Delta$, $0<\zeta<\pi$. If $\zeta$ is
commensurate with $\pi$ (in other words if $e^{i\zeta}$ is a root
of unity), then the integral in \eq{asympt} can be taken
explicitly in terms of $\psi$-function (logarithmic derivative of
$\Gamma$-function). In particular for $\zeta=\frac{\pi}{2}$ and
$\zeta=\frac{\pi}{3}$ (respectively $\Delta=0$ and $\Delta=1/2$)
we obtain from \eq{asympt}
\be{partcase}
\begin{array}{l}
{\dis \lim_{m\to\infty}\frac{\log\tau(m)}{m^2}= \frac{1}{2}\log2,
\qquad \Delta=0,}\non
{\dis \lim_{m\to\infty}\frac{\log\tau(m)}{m^2}=
\frac{3}{2}\log3-3\log2, \qquad \Delta=\frac{1}{2},}
\end{array}
\ee
which coincides with the known results obtained respectively in
\cite{ItsIKS93,ShiTN01,KitMST02c} and in \cite{RazS01,KitMST02b}.
For the particular case of the $XXX$ chain ($\Delta=1$, $\zeta=0$)
the asymptotic behavior can be evaluated also by the saddle point
method and it is given by
\be{XXXans}
\lim_{m\to\infty}\frac{\log\tau(m)}{m^2}=
\log\left(\frac{\Gamma(\frac{3}{4})\Gamma(\frac{1}{2})}
{\Gamma(\frac{1}{4})}\right)\approx\log(0.5991),
\ee
which is in good agreement with the known numerical result
$\log(0.598)$, obtained in \cite{BooKNS02}.

Below, we explain the main features of our method. A more detailed
account of the proofs and techniques involved will be published
later.

\section{The saddle point method}
The  multiple integral representation for $\tau(m)$ obtained in
\cite{KitMST02a} can be written in the form
\ba{EFPtautrans}
&&{\dis\hspace{-1mm}
\tau(m)=\left(\frac{i}{2\zeta\sin\zeta}\right)^m
\left(\frac{\pi}{\zeta}\right)^{\frac{m^2-m}{2}}
\int\limits_{\cal D} d^m\lambda\cdot F(\{\lambda\},m)}\non
&&{\dis\hspace{1mm}
\times\prod\limits_{a>b}^m\frac{\sinh\frac{\pi}{\zeta}
(\lambda_a-\lambda_b)}{\sinh(\lambda_a-\lambda_b-i\zeta)
\sinh(\lambda_a-\lambda_b+i\zeta)}
\prod\limits_{a=1}^m\left(
\frac{\sinh(\lambda_a-\frac{i\zeta}{2})\sinh(\lambda_a
+\frac{i\zeta}{2})}{\cosh\frac{\pi}{\zeta}\lambda_a}\right)^m
,}
\ea
with
\be{defF}
F(\{\lambda\},m)= \lim_{\xi_1,\dots\xi_m\to-\frac{i\zeta}2}
\frac{1}{\prod\limits_{a>b}^m\sinh(\xi_a-\xi_b)}
{\det}_m\left(\frac{-i\sin\zeta}
{\sinh(\lambda_j-\xi_k)\sinh(\lambda_j-\xi_k-i\zeta)}\right).
\ee
Here the integration domain ${\cal D}$ is $-\infty<\lambda_1
<\lambda_2<\cdots<\lambda_m<\infty$.

Following the standard arguments of the saddle point method we
estimate the integral \eq{EFPtautrans} by the maximal value of the
integrand. Let $\{\lambda'\}$ be the set of  parameters
corresponding to this maximum. They satisfy the saddle point
equations and for large $m$ we assume that their distribution can
be described by a density function $\rho(\lambda')$:
\be{dens}
\rho(\lambda'_j)=\lim_{m\to\infty}
\frac1{m(\lambda'_{j+1}-\lambda'_j)}.
\ee
Thus for large $m$, one can replace sums over the set
$\{\lambda'\}$ by integrals. Namely, if $f(\lambda)$ is integrable
on the real axis, then
\be{ruls}
\begin{array}{l}
{\dis\frac{1}{m}\sum_{j=1}^{m}f(\lambda'_j)\to
\int_{-\infty}^\infty f(\lambda)\rho(\lambda)\,d\lambda,}\non
{\dis\frac{1}{m}\sum_{j=1\atop{j\ne k}}^{m}
\frac{f(\lambda'_j)}{\lambda'_j-\lambda'_k}\to
V.P.\int_{-\infty}^\infty
\frac{f(\lambda)}{\lambda-\lambda'_k}\rho(\lambda)\,d\lambda,}
\end{array}
\qquad\qquad\qquad m\to\infty.
\ee
Due to \eq{ruls} it is easy to see that in the point $\lambda'_1,
\dots,\lambda'_m$ the products in the second line of
\eq{EFPtautrans} behave as $\exp(c\ m^2)$.

Our goal is now to estimate the behavior of the term
$F(\{\lambda'\},m)$. To do this we factorize the determinant in
\eq{defF} as follows for large $m$:
\ba{factor}
&&{\dis\hspace{1mm}
{\det}_m\left(\frac{-i\sin\zeta}
{\sinh(\lambda'_j-\xi_k)\sinh(\lambda'_j-\xi_k-i\zeta)}\right)}\non
&&{\dis\hspace{12mm}
=(-2\pi i)^m{\det}_m\Bigl(\delta_{jk}-\frac{K(\lambda'_j-\lambda'_k)}
{2\pi im\rho(\lambda'_k)}\Bigr)
{\det}_m\Bigl(\frac{i}{2\zeta\sinh\frac{\pi}{\zeta}
(\lambda'_j-\xi_k)}\Bigr),}
\ea
with
\be{Lieb}
K(\lambda)=\frac{i\sin2\zeta}{\sinh(\lambda-i\zeta)
\sinh(\lambda+i\zeta)}.
\ee
Indeed, for $m\to\infty$ one has
\ba{proddet}
&&{\dis\hspace{-5mm}
{\det}_m\Bigl(\delta_{jk}-\frac{K(\lambda'_j-\lambda'_k)}
{2\pi im\rho(\lambda'_k)}\Bigr)
{\det}_m\Bigl(\frac{i}{2\zeta\sinh\frac{\pi}{\zeta}
(\lambda'_j-\xi_k)}\Bigr)}\non
&&{\dis\hspace{-5mm}
={\det}_m\Bigl(\frac{i}{2\zeta\sinh\frac{\pi}{\zeta}
(\lambda'_j-\xi_k)}-\sum_{l=1}^{m}
\frac{K(\lambda'_j-\lambda'_l)}
{2\pi im\rho(\lambda'_l)}
\frac{i}{2\zeta\sinh\frac{\pi}{\zeta}
(\lambda'_l-\xi_k)}\Bigr)}\num
&&{\dis\hspace{1mm}
\longrightarrow
{\det}_m\Bigl(\frac{i}{2\zeta\sinh\frac{\pi}{\zeta}
(\lambda'_j-\xi_k)}-\int_{-\infty}^\infty
\frac{K(\lambda'_j-\mu)}
{2\pi i}
\frac{i\,d\mu}{2\zeta\sinh\frac{\pi}{\zeta}
(\mu-\xi_k)}\Bigr)}\non
&&{\dis\hspace{30mm}
=\left(\frac{1}{2\pi}\right)^m
{\det}_m\left(\frac{\sin\zeta}
{\sinh(\lambda'_j-\xi_k)\sinh(\lambda'_j-\xi_k-i\zeta)}\right)
.}\nonumber
\ea
Here we have used the fact that  the function
$i/2\zeta\sinh\frac{\pi}{\zeta}(\lambda_j-\xi)$ solves the Lieb
integral equation for the density of the ground state of the $XXZ$
magnet \cite{LieSM61} (and we have used the notations of
\cite{KitMST02a}) . The second determinant in the r.h.s. of
\eq{factor} is a Cauchy determinant, hence,
\be{newF}
F(\{\lambda'\},m)=(-i)^m\left(\frac{\pi}{\zeta}\right)^{
\frac{m^2+m}{2}}
\frac{\prod\limits_{a>b}^m\sinh\frac{\pi}{\zeta}
(\lambda'_a-\lambda'_b)}{\prod\limits_{a=1}^m
\cosh^m\frac{\pi}{\zeta}\lambda'_a}\cdot
{\det}_m\Bigl(\delta_{jk}-\frac{K(\lambda'_j-\lambda'_k)}
{2\pi im\rho(\lambda'_k)}\Bigr).
\ee
The behavior of the determinant in \eq{newF} can be estimated via
Hadamard inequality
\be{ham} |{\det}_m(a_{jk})|\le (\max |a_{jk}|)^m m^{\frac{m}{2}}.
\ee
applied to the above determinant and to the determinant of the
inverse matrix , which shows that
\be{limit}
\lim_{m\to\infty}\frac{1}{m^2}~\log{\det}_m
\left(\delta_{jk}-\frac{K(\lambda'_j-\lambda'_k)}
{2\pi im\rho(\lambda'_k)}\right)=0.
\ee
The last equation means that ${\det}_m\left(\delta_{jk}-
K(\lambda'_j-\lambda'_k)/ {2\pi im\rho(\lambda'_k)}\right)$ does
not contribute to the leading term of the asymptotics. Hence, it
can be excluded from our considerations.

Thus, up to subleading corrections of the exponential type the
emptiness formation probability behaves as
\be{form}
\tau(m)\longrightarrow\left(\frac{\pi}{\zeta}\right)^{m^2}
e^{m^2S_0},\qquad m\to\infty,
\ee
with
\ba{SS} &&{\dis\hspace{5mm}
S_0\equiv S(\{\lambda'\})=
\frac{1}{m^2}\sum_{a>b}^m\log\left(\frac{\sinh^2\frac{\pi}{\zeta}
(\lambda'_a-\lambda'_b)}{\sinh(\lambda'_a-\lambda'_b-i\zeta)
\sinh(\lambda'_a-\lambda'_b+i\zeta)}\right)}\non
&&{\dis\hspace{25mm} +\frac{1}{m}\sum_{a=1}^m\log\left(\frac
{\sinh(\lambda'_a-i\zeta/2) \sinh(\lambda'_a+i\zeta/2)}
{\cosh^2\frac{\pi}{\zeta}\lambda'_a}\right).}
\ea
Here the parameters $\{\lambda'\}$ are the solutions of the saddle
point equations
\be{sadp}
\frac{\partial S_0}{\partial\lambda'_j}=0.
\ee
In our case the system \eq{sadp} has the form
\ba{systsadp}
&&{\dis\hspace{1mm}
\frac{2\pi}{\zeta}\tanh\frac{\pi\lambda'_j}{\zeta}-
\coth(\lambda'_j-i\zeta/2)-\coth(\lambda'_j+i\zeta/2)}\non
&&{\dis\hspace{1mm} =\frac{1}{m}\sum_{k=1\atop{k\ne j}}^m\left(
\frac{2\pi}{\zeta}\coth\frac{\pi}{\zeta}(\lambda'_j-\lambda'_k)-
\coth(\lambda'_j-\lambda'_k-i\zeta)-\coth(\lambda'_j-\lambda'_k+i\zeta)
\right).}
\ea
Using \eq{ruls} we transform \eq{systsadp} into the integral equation
for the density $\rho(\lambda)$
\ba{inteq}
&&{\dis\hspace{-4mm}
\frac{2\pi}{\zeta}\tanh\frac{\pi\lambda}{\zeta}-
\coth(\lambda-i\zeta/2)-\coth(\lambda+i\zeta/2)}\non
&&{\dis\hspace{-4mm}
=V.P.\int_{-\infty}^\infty\left(
\frac{2\pi}{\zeta}\coth\frac{\pi}{\zeta}(\lambda-\mu)-
\coth(\lambda-\mu-i\zeta)-\coth(\lambda-\mu+i\zeta)
\right)\rho(\mu)\,d\mu.}
\ea
Respectively the action $S_0$ takes the form
\ba{action} &&{\dis\hspace{1mm} S_0=\int_{-\infty}^\infty
d\lambda \rho(\lambda) \log\left(\frac{\sinh(\lambda-i\zeta/2)
\sinh(\lambda+i\zeta/2)}
{\cosh^2\frac{\pi}{\zeta}\lambda}\right)}\non
&&{\dis\hspace{10mm} +\frac{1}{2} \int_{-\infty}^\infty
d\mu d\lambda \rho(\lambda)\rho(\mu)
\log\left(\frac{\sinh^2\frac{\pi}{\zeta}
(\lambda-\mu)}{\sinh(\lambda-\mu-i\zeta)
\sinh(\lambda-\mu+i\zeta)}\right).} \ea
Since the kernel of the integral operator in \eq{inteq} depends on
the difference of the arguments, this equation can be solved via
Fourier transform. Then
\be{Fsol}
\hat \rho(\omega)=\int_{-\infty}^\infty
e^{i\omega\lambda}\rho(\lambda)\,d\lambda=
\frac{\cosh\frac{\omega\zeta}{2}}
{\cosh\omega\zeta}.
\ee
Making the inverse Fourier transform we find
\be{sol} \rho(\lambda)=\frac{\cosh\frac{\pi\lambda}{2\zeta}}
{\zeta\sqrt{2}\cosh\frac{\pi\lambda}{\zeta}},
\ee
which obviously satisfies the needed normalisation condition for
density (integral on the real axis equals one). It remains to
substitute \eq{Fsol}, \eq{sol} into \eq{action}, and after
straightforward calculations we arrive at
\be{S0}
S_0=\frac{1}{2}\int\limits_{\mathbb{R}-i0}
\frac{d\omega}{\omega}\frac
{\sinh\frac{\omega}{2}(\pi-\zeta)\cosh^2\frac{\omega\zeta}{2}}
{\sinh\frac{\pi\omega}{2}\sinh\frac{\omega\zeta}{2}
\cosh\omega\zeta}.
\ee
Thus, we have obtained \eq{asympt}.

In  the case of the $XXX$ chain ($\Delta=1$) one should rescale
$\lambda_j\to\zeta\lambda_j$, $\xi_j\to\zeta\xi_j$ in the original
multiple integral representation \eq{EFPtautrans} for $\tau(m)$
and then proceed to the limit $\zeta\to0$ . The remaining
computations are then very similar to the ones described above,
therefore we present here only the main results. The behavior of
$\tau(m)$ is now given by
\be{XXXform}
\tau(m)\longrightarrow\pi^{m^2}
e^{m^2S_0},\qquad m\to\infty.
\ee
The action $S_0$ in the saddle point has the form
\ba{XXXact}
&&{\dis\hspace{-5mm}
S_0=\int_{-\infty}^\infty
\log\left(\frac{(\lambda-i/2)(\lambda+i/2)}
{\cosh^2\pi\lambda}\right)
\rho(\lambda)\,d\lambda}\non
&&{\dis\hspace{10mm} +\frac{1}{2} \int_{-\infty}^\infty
d\mu d\lambda \rho(\lambda)\rho(\mu)
\log\left(\frac{\sinh^2\pi(\lambda-\mu)}{(\lambda-\mu-i)
(\lambda-\mu+i)}\right).} \ea
The analog of the integral equation \eq{inteq} in the
$XXX$ case is
\be{XXXinteq}
2\pi\tanh\pi\lambda-\frac{2\lambda}{\lambda^2+\frac{1}{4}}
=V.P.\int_{-\infty}^\infty\left(
2\pi\coth\pi(\lambda-\mu)-
\frac{2(\lambda-\mu)}{(\lambda-\mu)^2+1}
\right)\rho(\mu)\,d\mu.
\ee
The solution of this equation is
\be{XXXrho}
\rho(\lambda)=\frac{\cosh\frac{\pi\lambda}{2}}
{\sqrt2\cosh\pi\lambda}.
\ee
Substituting \eq{XXXrho} into \eq{XXXact} we finally arrive at
\eq{XXXans}.

\section*{Acknowledgments}

N. K. is supported by  JSPS grant P01177. N. K. would like to
thank M. Jimbo for help. N. S. is supported by the grants
RFBR-02-01-00484, Foundation of the Support of Russian Science,
Leading Scientific Schools 00-15-96046, the Program Nonlinear
Dynamics and Solitons and by CNRS. J.M. M. is supported by CNRS.
V. T is supported by DOE grant DE-FG02-96ER40959 and by CNRS.  N.
K, N. S. and V. T. would like to thank the Theoretical Physics
group of the Laboratory of Physics at ENS Lyon for hospitality,
which makes this collaboration possible. We also would like to
thank the organizors of the "6th International Workshop Conformal
Field Theory and Integrable Models" held in Chernogolovka,
September 15-21, 2002, for the nice and stimulating scientific
(and extra-scientific) atmosphere they succeeded to generate.

\end{document}